\documentclass[12pt]{article}
\usepackage{jheppub}

\usepackage{amsfonts,amsopn,amsmath,amssymb}
\usepackage{amsthm}
\def\beq{\begin{equation}}  
\def\eeq{\end{equation}}  
\def\bea{\begin{eqnarray}}  
\def\eea{\end{eqnarray}}  
\def\half{\frac{1}{2}}  

\def\Tr{{\rm Tr}}

\newcommand{\RR}{{\mathbb R}}%Reals
\newcommand{\CC}{{\mathbb C}}%Complex
\newcommand{\ZZ}{{\mathbb Z}}%Integers
\newcommand{\HH}{{\mathbb H}}%quaternions
\newcommand{\CS}{\mathcal{S}}

\newcommand{\IR}{\mathbb{R}}
\newcommand{\IC}{\mathbb{C}}
\newcommand{\IZ}{\mathbb{Z}}

\def\Z{\mathbb{Z}}

\newcommand{\pa}{\partial}

\def\bea{\begin{eqnarray}}
\def\eea{\end{eqnarray}}
\def\be{\begin{equation}}
\def\ee{\end{equation}}
\def\ba{\begin{align}}
\def\ea{\end{align}}
\def\bse{\begin{subequations}}
\def\ese{\end{subequations}}

\newcommand{\bem}{\begin{pmatrix}}
\newcommand{\eem}{\end{pmatrix}}

\def\={\;  = \;}
\def\+{\, + \,}

\def\wt{\widetilde}
\def\wh{\widehat}
\def\bar{\overline}

\def\rt2{\sqrt{2}}

\def\vth{\vartheta}
\def\ve{\varepsilon}
\def\v{\varphi}

\def\g{\gamma}
\def\t{\tau}
\def\a{\alpha}
\def\b{\beta}

\def\p{\partial}
\def\tbar{\bar \tau}
\def\zbar{\bar z}
\def\ubar{\bar u}

\title{Moonshine in Fivebrane Spacetimes}

\preprint{NIKHEF2013-015, EFI-13-17}
\author[1]{Jeffrey A.~Harvey}
\author[2]{and Sameer Murthy}
\affiliation[1]{Enrico Fermi Institute and Department of Physics \\ 
University of Chicago \\
5620 Ellis Av., Chicago Illinois 60637, USA}
\affiliation[2]{Nikhef theory group, Science Park 105, \\
1098 XG Amsterdam, The Netherlands}
\abstract{
We consider type II superstring theory on $K3 \times S^1 \times \RR^{1,4}$ and study perturbative BPS states in 
the near-horizon background of two Neveu-Schwarz fivebranes 
whose world-volume wraps the $K3 \times S^1$ factor.  These states are counted by the spacetime helicity supertrace $\chi_2(\tau)$ which we evaluate. We find
a simple expression for $\chi_2(\tau)$ in terms of the completion of the mock modular form $H^{(2)}(\tau)$ that has appeared recently in studies of the decomposition of the elliptic genus of $K3$ surfaces into characters of the $N=4$ superconformal algebra and which manifests a moonshine connection to the Mathieu group~$M_{24}$. 
}

\begin{document}

\maketitle

\section{Introduction and motivation}

Mock modular forms  have appeared recently in a variety of physical and mathematical contexts. On the physical side, they
play a central role in the counting of black hole states in string theory \cite{DMZ} and in computations of the elliptic genus
of sigma models with non-compact target spaces \cite{TroostI,ESII,Ashok}. 
In a more mathematical direction, a particular mock modular form with $q$ expansion
\be \label{H2}
H^{(2)}(\tau) = \sum_{n=0}^\infty c^{(2)}(8n-1) \, q^{n-1/8}=2 q^{-1/8}(-1 +45 \, q + 231  \, q^2 + 770  \, q^3 + 2277 \, q^4 + \cdots )
\ee
appears in the decomposition of the elliptic genus of $K3$ surfaces into characters of the $N=4$ superconformal algebra
and reveals a mysterious moonshine property: the coefficients $45,231, 770, 2277$ are dimensions of irreducible representations of
the largest sporadic Matheiu group $M_{24}$ \cite{eot}.  This ``Mathieu Moonshine" has been further developed and tested through computation of the analogs of
the McKay-Thompson series of Monstrous Moonshine \cite{Thompson}, 
$H^{(2)}_g$ for $g \in M_{24}$ \cite{cheng,GaberdielI,GaberdielII,EguchiI}, and there
is now a proof \cite{Gannon} of the existence of an infinite-dimensional $M_{24}$--module
\be
K^{(2)}= \bigoplus_{n=0}^\infty K^{(2)}_{8n-1}
\ee
with ${\rm dim} K^{(2)}_{8n-1}= c^{(2)}(8n-1)$ for $n\geq 1$, although so far no explicit construction of such a module is known.

There are many reasons to expect a construction based on Conformal Field Theory (CFT). These include the fact that there is such a construction  \cite{FLMI,FLMII} 
that explains the similarly remarkable connection between the coefficients of the modular function $j(\tau)$ and dimensions of representations of the 
Monster group known as Monstrous Moonshine~\cite{conway_norton}, and also the properties of the $H^{(2)}(\tau)$ constructed by twisting by elements of~$M_{24}$.  Monstrous Moonshine appears to have a generalization dubbed {\it generalized Moonshine} by Norton \cite{norton} which involves the existence of
modular functions $Z_{g,h}(\tau)$ for congruence subgroups of $SL(2,\RR)$ for each pair of commuting elements $(g,h)$ in the Monster group. These were given
a conformal field theory interpretation in \cite{DGH} in terms of the partition function twisted by $h$ of an orbifold by $g$ of the Monster CFT.  A  construction of many
of these orbifold theories and their McKay-Thompson series can be found in \cite{Carnahan} and references cited therein.  Evidence for a  similar generalization
of Mathieu Moonshine has been presented in \cite{GPRV} and this can be regarded as further evidence that CFT is the correct framework in which to understand
Mathieu Moonshine. 

However, it is known that no classical~$K3$ surface
can exhibit the full~$M_{24}$ symmetry  \cite{Mukai, Kondo}. Furthermore, there is also good evidence that the superconformal field theory (SCFT) describing
any $K3$ surface also cannot exhibit the full $M_{24}$ symmetry \cite{Gaberdiel:2011fg}. Thus it seems likely  that one must look beyond the SCFT associated to $K3$
surfaces in the search for the origin of Mathieu Moonshine and an explicit construction of the infinite dimensional $M_{24}$ module $K^{(2)}$ (see however \cite{TW}
for an alternate point of view based on combining symmetry groups of distinct Kummer surfaces). 

Another clue is provided by the existence of
generalizations of the $(H^{(2)},M_{24}) $ moonshine  to an umbral moonshine for  vector-valued mock modular forms $H^{(X)}(\tau)$ and groups $G^{(X)}$ associated to
the 23 Niemeier lattices \cite{um,mum}.  While  some of the examples of umbral moonshine  can also be related to weight zero Jacobi forms, the Jacobi forms are not the elliptic genera of any compact Calabi-Yau manifold, and for other instances of  umbral moonshine it is weight one meromorphic Jacobi forms rather than weight zero Jacobi forms
that are the primary objects leading to vector-valued mock modular forms.  The existence of this large class of mock modular forms exhibiting Moonshine for finite groups but with no obvious connection to compact Calabi-Yau manifolds also points
towards the need for an extended notion of SCFT if there is to be a universal construction for the infinite dimensional modules suggested by these constructions.

Yet another clue for the origin of the $M_{24}$--module and its generalizations may lie in the following detail of the original 
observation of $M_{24}$ moonshine~\cite{eot}.  In order to obtain the mock modular 
form~\eqref{H2}, the term proportional to the massless character of the $N=4$ superconformal algebra had to be subtracted from the decomposition of 
the~$K3$ elliptic genus into $N=4$ characters. From the point of view of quantum field theory, removing
part of the spectrum of the theory generically violates some fundamental property like locality or a defining symmetry
of the theory. From this point of view, one may say that it is not too surprising that one does not find 
the~$M_{24}$--module in a simple direct manner in the $K3$ SCFT.  Such a module is more likely to be 
present in a theory whose full spectrum of BPS states is counted directly 
by the function $H^{(2)}$. Given the recent appearance of mock modular forms as the elliptic genera of 
non-compact CFTs, it would be particularly natural if the target space of the CFT that we are looking for 
involved both $K3$ and a non-compact space. 

The need to discard massless states is also reminiscent of the Frenkel-Lepowsky-Meurman construction of the  Monster module denoted by $V^\natural$ in \cite{FLMI,FLMII}.
In physics terminology the construction starts with the holomorphic part of the bosonic string on the torus $\RR^{24}/\Lambda_L$ where $\Lambda_L$ is the Leech lattice.
Since there are no points of length squared $2$ in $\Lambda_L$, this theory has $24$ massless states and a partition function that starts as
\be \label{monsterz}
Z(\tau) = q^{-1} + 24 + 196884 \, q + \cdots \,.
\ee
There is no $24$-dimensional irreducible representation of the Monster, and the construction of FLM proceeds by the construction of an asymmetric $\ZZ/2$ orbifold
which acts by $X^I \rightarrow - X^I$. This orbifold construction removes the $24$ massless states and does not produce any new massless states in the twisted sector
because the twist field has dimension $3/2$. The orbifold construction also preserves modular invariance and so leads to a partition function which differs from
Eqn.~\eqref{monsterz} only by the lack of a constant term.

Since mock modular forms appear in counting of supersymmetric, BPS black holes whose near horizon involves an Anti de Sitter space (AdS)  component, it
is also natural to wonder whether there might exist a BPS configuration of branes in string theory and an associated black hole counting problem where
$H^{(2)}$ and its generalizations might appear.
In light of the AdS/CFT correspondence this could provide a dual description of the  CFT's associated to Mathieu and Umbral Moonshine. This idea is also
supported by connections between semi-classical expansions in AdS and the Rademacher summability of \cite{Manschot:2007ha,Dijkgraaf:2000fq,DuncanFrenkel,ChengDuncanI,ChengDuncanII}.

As further motivation for the work presented here, we note that the appearance of  the mock modular form $H^{(2)}(\tau)$ in the decomposition of the elliptic genus of
$K3$ into characters of the $N=4$ superconformal characters is a worldsheet phenomenon.  It is often useful to find a translation of such worldsheet results into a spacetime
computation involving BPS states since in that context one can apply the powerful ideas of string duality. This translation between worldsheet and spacetime points of view
has been exploited heavily in the exact counting of BPS black hole states (see for example~\cite{Mandal:2010cj, Dabholkar:2010rm, DDMP}).
In the context
of Type II string theory on $K3 \times S^1$ or $K3 \times T^2$ one might naively expect the elliptic genus of $K3$ to count perturbative $1/4$ BPS states in intermediate
representations of the $N=4$ spacetime supersymmetry algebra since one can construct such states by combining purely left-moving excitations of the~$K3$ SCFT
with momentum and winding states on the~$S^1$ or~$T^2$. However it is known that these states in fact combine into long representations of the $N=4$ supersymmetry
algebra and so do not contribute to the spacetime helicity index that counts BPS states \cite{Kiritsis}.  Thus to find some spacetime, BPS image of the worldsheet decomposition
it is natural to look at systems with the equivalent of $N=2$ spacetime supersymmetry rather than $N=4$ supersymmetry since in that case it is known that there are BPS states which are counted
by the elliptic genus of $K3$. For example, this can be seen in the computation of threshold corrections in $N=2$ heterotic string compactifications in \cite{HarveyMoore} which depend on the new supersymmetric index which in turn can be seen to count the difference
between BPS vector and hypermultiplets. Connections between Mathieu Moonshine and threshold corrections in $N=2$ heterotic string compactifications  and their type
II duals were recently explored in \cite{Cheng:2013kpa}.

In this paper we take a first step in this direction through the computation of the second helicity index 
(often called the BPS index) $\chi_2(\tau)$ in the near horizon geometry of a background of two Neveu-Schwarz fivebranes in type II string theory on $K3 \times S^1$.  This background has a spacetime supersymmetry algebra which has the same number of supersymmetries as an $N=2$ theory in $\RR^{1,3}$ and has perturbative BPS states which are counted by the index $\chi_2(\tau)$. We find that $\chi_2(\tau)= - (1/2)  \,  \eta(\tau)^{3} \widehat H^{(2)}(\tau)$ where $\eta(\tau)$ is the
Dedekind eta function and $\widehat H^{(2)}(\tau)$ is the completion of the mock modular form $H^{(2)}(\tau)$ determined by its shadow
$g(\tau) = 24 \, \eta(\tau)^{3}$.  
The outline of this paper is as follows.  In the second
section we discuss the fivebrane background we utilize and some details of the underlying conformal field theory. The third section goes through the
calculation and interpretation of the BPS index while the fourth section discusses some properties of mock modular forms and the modification to this computation of the  BPS index when we twist the theory by symplectic automorphisms
of the $K3$ surface. The final section offers conclusions and a discussion of interesting directions suggested by our results.   Some details of the analysis
of an integral first analyzed by Gaiotto and Zagier are presented in Appendix A while Appendix B summarizes our conventions for theta functions as well as some
Riemann theta relations that are used in our computations.

\section{Wrapped fivebranes and the $K3 \times SL(2,\RR)/U(1)$ SCFT \label{superstring}}

Consider type II string theory in the background of $k$ NS5-branes in ten-dimensional flat space. 
In the RNS formalism, fundamental string propagation in the near-horizon region of the branes is described by a two-dimensional superconformal field theory \cite{Callan:1991dj}, which we denote as:
\be \label{CHS}
\IR^{1,5} \times \rho \times SU(2)_{k} \, . 
\ee
Here  the first factor corresponds to the space-time which the 5-branes span, and represents six free bosons 
as well as their $N=1$ superpartners. The second factor corresponds to an $N=1$ linear dilaton 
theory with slope\footnote{We will set $\a'=2$ throughout this paper.} $Q=\sqrt{\frac2k}$ and central charge 
$c= \frac32 + 3 Q^{2}$, and represents the radial direction in the $\IR^{4}$ 
transverse to the branes. The third factor is an $N=1$~$SU(2)$ WZW model at level $k$ with central charge 
$c=\frac92 - \frac6k$, and represents the~$S^{3}$ of the transverse~space. 

To make a consistent string theory one must introduce the $N=1$ ghost system $(b,c,\b,\g)$ with central 
charge $c=-15$. Spacetime supersymmetry can be introduced by the usual method of identifying 
an~$N=2$ structure in the above SCFT, and by imposing the GSO projection. 
This gives us a theory with 8 left-moving and 8 right-moving supercharges which transform 
non-trivially under the~$SU(2)_{L} \times SU(2)_{R} = SO(4)$ rotations of the transverse $\IR^{4}$.

The string coupling is given in terms of the radial coordinate by  $g_{s} = g^{(0)}_{s}e^{-\rho}$ so that fundamental strings are weakly coupled 
in the asymptotic region $\rho \to \infty$, and they become arbitrarily strongly coupled deep inside the throat of 
the branes at $\rho \to -\infty$. In order to study string perturbation theory  we would like to cap off the 
strong-coupling singularity. A way of doing so was suggested in \cite{Giveon:1999px}, by spreading out the 
5-branes on a ring in the transverse $\IR^{4}$ thus breaking the $SO(4)$ R-symmetry to $U(1) \times \ZZ/k$. 
The authors of \cite{Giveon:1999px} proposed that the SCFT corresponding to this configuration is:
\be \label{GKP}
\IR^{1,5} \times \Big( \frac{SL(2,\IR)_{k}}{U(1)} \times \frac{SU(2)_{k}}{U(1)} \Big) \Big/ (\ZZ/k) \, ,
\ee
where the $\ZZ/k$ orbifold is required to implement the integrality of charges on which a~$\ZZ/2$ GSO projection \cite{Israel:2004ir} can act. 
The level indicated in both the WZW models is the supersymmetric level, and the levels 
of the two bosonic algebras are related to $k$ as 
\be\label{kkbrel}
k^{sl(2)}_{B} \= k+2 \ , \qquad k^{su(2)}_{B} \= k-2 \, .
\ee 

The $\frac{SU(2)_{k}}{U(1)}$ factor in \eqref{GKP} is the well-understood compact $N=2$ coset of 
central charge~$c=3-\frac6k$. 
The $\frac{SL(2,\IR)_{k}}{U(1)}$ factor in \eqref{GKP} denotes the non-compact coset theory called the cigar theory or 
the Euclidean black hole \cite{Witten:1991yr}, with $c=3+\frac6k$. In the large $k$ limit, the coset has a geometric 
picture as a sigma model on the cigar geometry with curvature proportional to $1/k$. The algebraic 
approach, on the other hand, is exact in~$k$. 
For the purposes of computing Euclidean path-integrals, we follow the treatment of \cite{Gawedzki:1991yu, GK, Karabali, Schnitzer}), 
in which the cigar theory is defined as the Euclidean coset $H_{3}^{+}/U(1)$ with 
$H_{3}^{+} = SL(2,\IC)/SU(2)$.

Asymptotically, the cigar model consists of a linear dilaton direction $\rho$ with 
slope $Q=\sqrt{\frac{2}{k}}$, and a $U(1)$ direction $\theta$ with $ \theta \sim \theta +\frac{4\pi}{Q}$, 
and two fermions $(\psi_{\rho}, \psi_{\theta})$. 
Together, they make up an $N=2$ SCFT with the following holomorphic currents (see e.g.~\cite{Murthy:2003es}):
\bea\label{ntwowss}
 T_{\rm cig} & \= & -\half (\partial \rho)^2 -\half (\partial\theta)^2 - \half
 (\psi_\rho\partial \psi_\rho + \psi_\theta \partial \psi_\theta)
 - \half Q \partial^2\rho \, , \cr
 G^\pm_{\rm cig} & \= &\frac{i}{2} (\psi_\rho \pm i\psi_\theta)\partial(\rho \mp
 i\theta) +\frac{i}{2} Q\partial (\psi_\rho \pm i\psi_\theta) \, , \cr
 J_{\rm cig} & \= & -i\psi_\rho\psi_\theta +iQ\partial \theta \, ,
\eea
as well as their anti-holomorphic counterparts. 
In combination with the $SU(2)/U(1)$ coset and the flat directions, one recovers the theory~\eqref{CHS}
in the asymptotic region. The strong coupling region, however, has now been capped off by the geometry 
of the cigar, and the string coupling has a maximum at the tip of the cigar, the value of which is a modulus of the 
string theory. 

The full $N=2$ worldsheet currents of the theory include the currents coming from the flat space and 
$SU(2)/U(1)$ factors in \eqref{GKP}. Using this $N=2$ structure, we can now construct spin fields and spacetime 
supersymmetry. We have 8~left-moving and 8~right-moving spacetime supercharges $\CS_{\a}, \wt \CS_{\a},$ 
that obey the algebra
\be\label{susyalg}
\{\CS_\alpha,\bar \CS_\beta\} \= 2\gamma^\mu_{\alpha\beta}P_\mu \, , \qquad 
\{\wt \CS_\alpha,\bar{\wt \CS}_\beta\} \= 2\gamma^\mu_{\alpha\beta}P_\mu \,  , \qquad \qquad \mu = 0,1\cdots 5 \, .
\ee
The spinors~$S_{\a}$ are minimal Weyl spinors of~$Spin(1,5)$, and the bar denotes charge conjugation. 
In the~IIA theory, the chirality of the left-movers and 
the right-movers are the same, while in the~IIB theory they are opposite.

We also have a global $U(1)$ symmetry coming from the momentum around the circle~$\theta$:
\be \label{Jsp}
J_{\rm sp} \= P_{L}^{\theta} + P_{R}^{\theta}
\, \equiv \, \frac{i}{Q} \oint \p \theta \, dz \+ \frac{i}{Q} \oint \overline \p \wt \theta  \, d\bar z\, , 
\ee
under which all the spacetime supercharges are charged:
\be \label{Rsp}
  [J_{\rm sp}, \CS_\alpha]\=\half \CS_\alpha \, , \qquad 
  [J_{\rm sp}, {\bar \CS}_{\alpha}]\=-\half {\bar \CS}_{\alpha} \, . 
\ee
There is a similar expression for the right-moving supercharges.
The $U(1)$ momentum symmetry is thus a spacetime R-symmetry and the  
spacetime fermion number is $(-1)^{F_{\rm s}} = e^{2 \pi i J_{sp}}$.

It is clear from the above worldsheet construction that in order to study NS5-branes wrapped on a $K3$ surface,
one simply replaces the $\IR^{1,5}$ by $\IR^{1,1} \times K3$. In this case the~$K3$ breaks a further half of the 
supersymmetry, and we get a superstring theory with 4 left-moving and 4 right-moving supercharges. 
Translation invariance along the $K3$ directions is now broken, and the supercharges anti-commute to 
translations along the~$\IR^{1,1}$ directions.

At level $k=2$, when the model represents the theory with two NS5-branes, something special 
happens\footnote{The theory \eqref{GKP} at $k=2$ is also the end-point $d=6$ of another family of interesting 
superstring theories called non-critical superstrings~\cite{Kutasov:1990ua}, defined as an $N=2$
generalization of Liouville theory combined with $d$ flat spacetime dimensions. It was shown 
in~\cite{Hori:2001ax} that the $N=2$ Liouville theory is indeed mirror symmetric to the cigar 
supercoset.}~\cite{Murthy:2003es}. 
The compact coset $SU(2)_{k}/U(1)$ (with central charge~$c=3-6/k$) disappears, and the 
free boson $\theta$ is equivalent to two free-fermions. These two fermions combined with the 
fermion $\psi_{\theta}$ obey an $SU(2)$ algebra, and these enhanced symmetries give rise to the 
expected $SU(2)_{L} \times SU(2)_{R}$ symmetries of the CHS model \eqref{CHS}. 
On separating the two five-branes in the transverse $\IR^{4}$ this is broken to an 
$SU(2) \times (\ZZ/2)$ global symmetry (instead of~$U(1) \times (\ZZ/k)$ for $k >2$), as expected 
from the spacetime picture of two 5-branes.

Finally we can, without any further issues, consider the single flat spatial direction to be a large circle to get 
type II superstring theory on 
\be \label{wsSCFT}
\IR_{t} \times S^{1} \times K3 \times \Big( \frac{SL_{2}(\IR)_{k=2}}{U(1)} \Big)\Big/ (\ZZ/2) \, , 
\ee
which is the model we shall study in this paper.

\subsection{The generating function of perturbative BPS states}

We would like to study the degeneracies of perturbative BPS states in the string theory~\eqref{wsSCFT}. 
We consider a fundamental type II string propagating in time and wrapping the circle in~\eqref{wsSCFT}. 
The covariant RNS description of the string has oscillators associated with the $\IR_{t} \times S^{1}$ 
directions which are cancelled in all physical computations by the oscillators of the $(b,c,\b,\g)$ superghost 
system that gauge the $N=1$ supergravity on the string world-sheet. 
One can also directly choose a gauge condition
on the string world-sheet that eliminates the unphysical oscillators in the $\IR_{t} \times S^{1}$ directions. 
To this end one can make a small modification to the usual light-cone gauge condition in $\IR^{1,1}$ so 
as to keep only the transverse oscillators on the string world-sheet \cite{DabMur}. This leaves us with an $N=(4,4)$ 2d SCFT with central charge~$c= \wt c =12$ described by 
\be \label{wsSCFTlc}
K3 \times \Big( \frac{SL_{2}(\IR)_{k=2}}{U(1)} \Big)\Big/ (\ZZ/2) \, .
\ee

If the string has momentum and winding
labelled by integers $n,w$ respectively, and we choose $n \ge 0, w \ge 0$, then in this compact 
light-cone gauge we have 
\be \label{levelmatch}
M^2  \= \frac{q_R^2}{2} + \tilde h+a \= \frac{q_L^2}{2} + h+a \, , 
\ee
where $M \equiv |p_{0}|$ denotes the energy of a state corresponding to an excitation of the SCFT~\eqref{wsSCFTlc} with left and right-moving conformal 
weights~$h, \tilde h$ and with
\be \label{qlrdef}
q_{R,L} \= \frac{n}{R} \pm \frac{wR}{2}
\ee
where $R$ is the radius of the $S^1$. The constant in Eq.~\eqref{levelmatch} arises from the zero point 
energy and is equal to $a =-\frac12$.

%If the string has momentum and winding
%labelled by integers $n,w$ respectively, and we choose $n \ge 0, w \ge 0$, then in this compact 
%light-cone gauge we have 
%\be \label{levelmatch}
%M^2  \= \frac{q_R^2}{2} + \tilde h-\frac12 \= \frac{q_L^2}{2} + h-\frac12 \, , 
%\ee
%where $M \equiv |p_{0}|$ denotes the energy of a state corresponding to an excitation of the SCFT~\eqref{wsSCFTlc} with left and right-moving conformal 
%weights~$h, \tilde h$ and with
%\be \label{qlrdef}
%q_{R,L} \= \frac{n}{R} \pm \frac{wR}{2}
%\ee
%where $R$ is the radius of the $S^1$. 

From the asymptotic supersymmetry algebra~\eqref{susyalg} compactified on~$K3$, it follows that states 
annihilated by the right moving supercharges have~$M=|q_{R}|$ which implies that~$\tilde h+a_R=0$.  
For such states, the level-matching condition~\eqref{levelmatch} implies that the product of the winding and 
momenta 
\be \label{nw}
nw \= h+a_L \, . 
\ee

Perturbative BPS states in string theory in flat space can be summarised in a succinct way in terms of 
spacetime helicity supertraces~\cite{Kiritsis}. We would like to compute similar BPS indices for our string theory. 
In particular, we are interested in generating functions of the form
\be \label{defchin}
\chi_{n}(\t) \= \Tr \, (J_{\rm sp})^{n} \, (-1)^{F_{\rm s}} \, q^{L_{0}- c/24} \, \bar q^{\wt L_{0}-\wt c/24} \,  \quad \qquad q = e^{2 \pi i \t} \, , 
\ee
where $\t$ is the modular parameter of the world-sheet torus and $\Tr$ indicates a sum over all the states in 
the theory~\eqref{wsSCFTlc}.
In the RNS formalism it represents a sum over Ramond and Neveu-Schwarz (NS)
sectors with chiral GSO projections.  

Our general strategy to obtain $\chi_n(\t)$ is to first compute 
\be
\chi(\t, z) \= \Tr \, (-1)^{F_{\rm s}} \, q^{L_{0}-c/24} \, \bar q^{\wt L_{0}-\wt c/24} \, \zeta^{P_{L}^{\theta}} 
\, \bar \zeta^{P_{R}^\theta}  \, , 
\qquad \quad  \zeta = e^{2 \pi i z} \, ,
\ee
and then act on it by the operator $\big(\frac{1}{2 \pi i}(\p_{z} - \p_{\bar z})\big)^{n} \big|_{z=\zbar=0}$. 

In a theory with $N=2$ spacetime supersymmetry in four dimensions, the quantity~$\chi_{0}(\t)$ receives 
a vanishing contribution from long as well as short multiplets in the theory~\cite{Kiritsis}, this turns out to be 
true for our situation as well. 
We shall focus on the first non-vanishing helicity supertrace $\chi_{2}(\t)$ here.

Our computation has both a space-time and a world-sheet interpretation.  In the space-time without NS5-branes
the partition functions~\eqref{defchin} 
(after adding in the partition function of the winding and momentum modes around the~$S^{1}$) would be 
precisely the Euclidean version of the helicity supertraces in four dimensional string theory on $K3 \times T^{2}$, 
as computed say in~\cite{Kiritsis}. Indeed, one can check that the operator $J_{0}$ is the charge of the $U(1)$ that 
rotates two directions in the~$\IR^{4}$ transverse to the 5-branes~\cite{Giveon:1999px}. 

We generalize this counting by working in a background sourced by two heavy defects, the NS5-branes. The first non-zero BPS index
is then $\chi_2(\tau)$ and from~ \eqref{nw}, we see that the coefficients of the generating 
function are  the degeneracies of such states in terms of the T-duality charge invariant~$nw$~\cite{Dabholkar:1989jt, Dabholkar:1990yf}. 
More precisely, we should sum over the partition function associated to the momentum and winding states in computing the full BPS index of the
theory leading to
\be \label{compsum}
\sum_{n,w \in \ZZ} q^{q_L^2/2} \, {\bar q}^{q_R^2/2} \, \chi_2(\tau)
\ee
with $q_{L,R}$ given in \eqref{qlrdef}. We will see that $\chi_2(\tau)$ is not holomorphic, but has a holomorphic part given by
\be \label{defcN}
\chi_2(\tau) \big|_{hol} = - \frac{1}{2} \, \eta(\tau)^3 H^{(2)}(\tau) = \sum_{N=0}^\infty c(N) \, q^N 
\ee
which we will interpret as counting $1/4$ BPS states that are localized near the tip of the cigar. The physical states satisfying level-matching are then
those with equal powers of $q$ and $\bar q$ in \eqref{compsum}, that is those states with
\be
N = \frac{1}{2} (q_R^2-q_L^2) = nw \, .
\ee
We can thus interpret the coefficients $c(N)$ as counting the contribution of $1/4$ BPS states to the BPS index in the near horizon geometry of two NS5-branes
with mass squared $M^2=q_R^2/2$ and with T-duality invariant $nw$ equal to $N$. 

% ZZZ New here.
We expect to find a relation between the coefficients~$c(N)$ in~\eqref{defcN} and 
the degeneracy of small BPS black holes with charges~$(n,w)$ in the background of two NS5-branes. 
These black holes have vanishing horizon area in the two-derivative gravitational theory, but in a similar situation
in flat space they can gain a finite string-scale size upon introducing higher-derivative corrections~\cite{Sen:2004dp}. 
Since the function~$\chi_{2}(\t)|_{hol}$ does not have a polar term in its~$q$-expansion, the coefficients~$c(N)$ 
do not grow exponentially in~$\sqrt{N}$ as~$N\to \infty$ as one might expect from the black hole picture. 
Perhaps the details of the relation between the gravitational index and degeneracy~\cite{Dabholkar:2010rm} 
plays a role in resolving this puzzle.

\section{Computation of the BPS index \label{compBPS}}

In this section we enter into the details of the computation of the BPS index. The reader who is only interested in the final answer can skip ahead to
~\eqref{chi2int}.  Before getting started we note two general features of the analysis. 
First, in the RNS formulation, the two factors in the SCFT~\eqref{wsSCFTlc} are essentially decoupled except that the sum over 
the different fermion periodicities ties together the various free field pieces in the 
partition function. We shall use the description of~$K3$ as a~$T^{4}/ (\ZZ/2)$ orbifold, but as we shall see, 
the final answer depends only on the elliptic genus of $K3$ which 
is invariant across the $K3$ moduli space. Second, 
the partition function of the $SL(2,\IR)/U(1)$ coset involves an integral over  a gauge field zero mode which is the source of
the integral over the variable $u$ in ~\eqref{chi2int}. 

The analysis involves a number of Jacobi theta functions. 
Our conventions for these as well as some useful identities they obey are given in Appendix~\ref{thetaids}. 

We now describe the relevant partition functions of the various pieces that make up the SCFT~\eqref{wsSCFTlc}.
In the fermionic sector we present the NS sector partition functions explicitly. The partition functions in the other sectors $\rm NS (-1)^{F}, R, R(-1)^{F}$ follow easily from the free fermion analysis, one can also write them using worldsheet $N=2$ spectral flow applied to the NS partition function.

\subsection{The cigar piece}

The functional integral for the indexed partition function of the $SL(2,\IR)_{k}/U(1)$ (cigar) SCFT has recently been explicitly computed in \cite{TroostI,ESII,Ashok}\footnote{The holomorphic part of this partition function had been presented earlier in \cite{ES04}.} based on the work of \cite{Gawedzki:1991yu,Karabali,Schnitzer}. 
We shall follow this treatment in what follows. The main idea is to express the~$G/H$ WZW coset as~$G \times H^{\IC}/H$ 
where~$H^{\IC}$ is a complexification of the subgroup~$H$ that is gauged. To this one adds a~$(b,c)$ ghost 
system of central charge~$c=-\dim(H)$. The three pieces are coupled only via zero modes. 

Our case of interest here is the supersymmetric $SL(2,\IR)/U(1)$ WZW coset. 
The theory has a bosonic~$H_{3}^{+}$ WZW model at level $k+2$ of which a~$U(1)$ subgroup is gauged, and 
two free fermions $\psi^{\pm}$ (and their right-moving counterparts). 
The coset~$H^{\IC}/H$ is represented by the compact boson~$Y$. The zero mode in question is the holonomy of the gauge field around the two cycles of the torus which is represented by a complex parameter\footnote{Throughout this paper, we will use the subscripts 1 and 2 on a complex variable to denote its real and imaginary parts, i.e.~$\t=\t_{1}+i \t_{2}$, $u=u_{1}+i u_{2}$ etc.} $u = a \t +b$. The~$(b,c)^{\rm cig}$ ghost system has central charge $c=-2$. 
The bosonic~$SL(2,\IR)$, the two fermions, the~$Y$ boson, and the~$(b,c)$ ghosts are all solvable theories and are 
coupled by the holonomy~$u$ that has to be integrated over the elliptic curve $E(\t)=\IC/(\IZ \tau + \IZ)$.

The various pieces have the following contributions. 
The bosonic $H_{3}^{+} = SL(2,\IC)/SU(2)$ model contributes:
\be \label{H3plus}
Z_{H_{3}^{+}}(\t,u) \= \frac{(k+2) \sqrt{k}}{\t_{2}^{1/2}} \, e^{{2 \pi u_{2}^{2} / \tau_{2}}} \,  \frac{1}{|\vartheta_{11}(\tau,u)|^{2}} \, . 
\ee
The $(b,c)^{\rm cig}$ ghosts have the contribution:
\be\label{bcpartfn}
Z_{\rm gh}(\t) \= \t_{2} \, |\eta(\t)^{2} |^{2} \ . 
\ee
The two left-moving fermions $\psi^{\pm}$ have a contribution in the NS sector\footnote{The prefactor in front of the 
usual expression for free fermions arises because of a factor of $k+2$ in the action of these fermions. This prefactor 
cancels an equivalent one in the numerator of the bosons in~\eqref{H3plus}.}  \cite{AGMV}: 
\be\label{psipm}
Z^{\rm NS}_{\rm \psi^{\pm}}(\t) \=  \frac1{\sqrt{k+2}} \,e^{{-\pi u_{2}^{2} / \tau_{2}}} \, e^{{2 \pi i u_{1} u_{2} / \tau_{2}}} \, 
\frac{\vartheta_{00}(\tau,u)}{\eta(\t)} \, ,
\ee
and their right-moving counterparts have a similar contribution:
\be\label{psipmbar}
Z^{\rm NS}_{\rm \wt \psi^\pm}(\t) \= \frac1{\sqrt{k+2}} \, e^{{-\pi u_{2}^{2} / \tau_{2}}} \, e^{{-2 \pi i u_{1} u_{2} / \tau_{2}}} \, 
\frac{\overline{\vartheta_{00}(\tau, u)}}{\overline{\eta(\t)}} \, .
\ee

Now we come to the boson $U(1)_{Y}$. The matching to the asymptotic fields \eqref{ntwowss}
shows that $\psi^{\pm} = \psi^{\rho} \pm i \psi^{\theta}$, and the boson $Y^{u} \equiv Y + \Phi[u]$ with 
$\Phi[u] = \frac{i}{\t_{2}} (w\bar u - \bar w u)$ should be identified with the boson $\theta$. 
(The notations are those of \cite{Ashok}.)
For the case~$k=2$, we know that the boson $\theta$ is equivalent to two free fermions $\chi^{\pm}$, so that in the 
asymptotic region the variables are the fields\footnote{The fields~$(\rho, \psi_{\rho})$ form an $N=1$ 
theory, and the three free fermions~$(\psi^{\theta}, \psi^{1,2})$ form an $N=1$ $SU(2)$ current algebra 
at level~$k=2$. This~$SU(2)$ and the corresponding one from right-movers form the currents of the 
asymptotic~$SO(4)$ theory of the theory of two 5-branes. This~$SO(4)$ is then broken 
to~$SU(2) \times \IZ_{2}$ by the cigar interactions, see \cite{Murthy:2003es}, \S3.4 for details.}
$(\rho, \psi_{\rho}, \psi^{\theta}, \chi^{\pm})$. These four fermions along with the four fermions of $K3$
and the two fermions of $\IR_{t} \times S^{1}$ are the analog of the ten free fermions of type II string 
theory in flat space. 
These considerations suggest that the boson $Y^{u}$ should 
really be treated as a pair of fermions with their corresponding spin structure. 
The same conclusion can also be reached by looking at the worldsheet $N=2$ algebra \eqref{ntwowss}
which is used to build spacetime supercharges.

The boson $Y^{u}$ is translationally charged under the potential $u$ (see eqn.~(2.21) of \cite{ESII}),
and this means that the fermions $\chi^{\pm}$ have charges~$\pm 1$ under the corresponding~$U(1)$ current. 
The contribution of these fermions is: 
\be\label{psi12}
Z^{\rm NS}_{\rm \chi^{\pm}}(\t) \=  e^{{-\pi (u_{2}+z_{2})^{2} / \tau_{2}}} \, 
e^{{2 \pi i (u_{1} +z_{1})(u_{2} +z_{2})/ \tau_{2}}} \,  \frac{\vartheta_{00}(\tau,z+u)}{\eta(\t)} \, ,
\ee
and their right moving counterparts contribute:
\be\label{psi12bar}
Z^{\rm NS}_{\rm \wt \chi^\pm}(\t) \=  e^{{-\pi (u_{2}-z_{2})^{2} / \tau_{2}}} \,  
e^{{-2 \pi i (z_{1} -u_{1})(z_{2} -u_{2})/ \tau_{2}}} \, \frac{\overline{\vartheta_{00}(\tau, z - u)}}{\overline{\eta(\t)}} \, .
\ee
We see here that the left- and right-movers are charged oppositely under the $U(1)$ gauge field -- 
this can be traced to the fact 
that the coset is an axial gauging of the $H_{3}^{+}$ WZW model\footnote{One can compare the relative charge 
assignments of the boson $Y^{u}$ with respect to the momentum $U(1)$ ($\p Y$) and the the gauged $U(1)$ ($u$). 
This is written down clearly in \cite{Ashok}, equations (2.28)--(2.32). We see that, indeed, the charge assignments are consistent
with the assignment of the potentials in~\eqref{psi12},~\eqref{psi12bar}.}.

\subsection{The $K3$ piece}

We evaluate the $K3$ partition function at an orbifold point $T^{4}/ (\ZZ/2)$. 
The $T^{4}$ SCFT consists of four bosons $X^{i}$ and four fermions $\xi^{i}$, $i=1,\cdots,4$.
The $\ZZ/2$ orbifold acts by reflection through the origin on the four bosons (i.e.~as $X^{i} \to - X^{i}$).
Supersymmetry requires that the orbifold acts in exactly the same way on the four 
fermions (i.e.~as $\xi^{i} \to - \xi^{i}$) .  

Following standard procedure for orbifold theories, we need to sum over the twisted sectors
and project to $\ZZ/2$ invariant states. Denoting the $\ZZ/2$ valued twist by $r \in \{0,1 \}$, this sum 
is equivalent to summing over all possible periodicities in both the directions of the worldsheet torus, 
i.e.~over the sectors $(r,s), r,s=0,1$. 

The partition function of the bosons in the untwisted sector is given by 
\be
Z^{\rm bos}_{K3 \, (0,0)}(\t) = \frac{\Theta^{4,4}(\t ,\tbar)}{|\eta(\t)^{4}|^{2}} \, ,
\ee
where the $\Theta^{4,4}$ indicates the sum over the $\Gamma^{4,4}$ Narain lattice of the $T^{4}$.
The left moving fermionic oscillator modes (with NS boundary conditions) is:
\be\label{K3untw}
Z^{\rm fer \, NS}_{K3 \, (0,0)}(\t) =  
\frac{\vartheta_{00}(\t,0)^{2}}{\eta(\t)^{2}} \, ,
\ee
and there is a corresponding factor from the right movers. Note that the fields of the $K3$ are not charged 
under the chemical potentials $u$ (from the gauging of the coset), nor are they charged under the 
spacetime $U(1)$ R-symmetry.

In the sectors $(r,s) \neq (0,0)$, there is no lattice sum. The bosonic partition function of the oscillator 
modes is:
\be
Z^{\rm bos}_{K3 \, (r,s)}(\t) = 16 \, \bigg| \frac{\eta(\t)^{2}}{\vartheta_{11}(\t,(s+r\t)/2)^{2}} \bigg|^{2} \, .
\ee
The left-moving NS sector fermionic partition function is:
\be\label{K3rstw}
Z^{K3\, \rm NS}_{(r,s)}(\t) =  \frac{\vartheta_{00}(\t,(s+r\t)/2)^{2}}{\eta(\t)^{2}} \, ,
\ee
and there is a corresponding partition function for the right-movers. 

\subsection{Putting the pieces together}

The full partition function is obtained by multiplying the various bosonic and fermionic pieces of the cigar 
and the $K3$ SCFT, summing over $\rm NS, NS(-1)^{F}$, $\rm R$ and $\rm R(-1)^{F}$ fermion periodicities 
in each $(r,s)$ twisted sector, and then summing over the twists. We include a factor of $1/2$ for each projection in the sum.

\paragraph{The untwisted sector}

In the untwisted sector, we obtain:
\bea\label{untwpartfn}
Z_{(0,0)}(\t , \tbar, u, \ubar, z, \zbar) = && \frac{1}{\sqrt{2}} \, 
\t_{2}^{1/2} \, e^{{-2 \pi u_{2}^{2} / \tau_{2}-2 \pi z_{2}^{2} / \tau_{2}}} \,
e^{{4 \pi i (u_{1}z_{2}+z_{2}u_{1})/ \tau_{2}}} \, \times \\
&&  \qquad \times \, \frac{ |\eta(\t)^{2} |^{2} }{|\vartheta_{1}(\tau,u)|^{2}} \, \frac{\Theta^{4,4}(\t ,\tbar)}{|\eta(\t)^{4}|^{2}} \, Z^{\rm fer, sum}_{(0,0)}(\tau, u, z) \, 
 \bar Z^{\rm fer, sum}_{(0,0)}(\tbar, \ubar, \zbar) \, , \nonumber
\eea
where $Z^{\rm fer, sum}_{(0,0)}(\t,u,z)$ denotes the sum over all the left-moving fermionic pieces of the theory, 
and is given by:
\bea\label{untwpartfnfull}
Z^{\rm fer, sum}_{(0,0)}(\tau, u, z)  & = & \half 
\frac{1}{\eta(\t)^{4}}\bigl( \vartheta_{00}(\tau, u) \vartheta_{00} \left( \tau, z+u \right)   \vartheta_{00}(\tau)^{2}
-  \vartheta_{01}(\tau, u)  \vartheta_{01} \left( \tau, z+u \right) \vartheta_{01}(\tau)^{2}
\cr
&  & \qquad \qquad 
 -  \vartheta_{10} ( \tau,u)  \vartheta_{10} \left( \tau, z+u \right)  \vartheta_{10}(\tau)^{2}
-  \vartheta_{11}(\tau, u)  \vartheta_{11} \left( \tau, z+u \right)  \vartheta_{11}(\tau)^{2}  \bigr) \, ,  \cr
& = & \half  \frac{1}{\eta(\t)^{4}} \, \vartheta_{11}(\tau,  z/2)^{2}  \, 
\vartheta_{11}( \tau, z/2+u)^{2}   \, .
\eea
In going to the second line, we have used the Riemann identity R5 of \cite{Mumford}. 
Similarly, the right-movers evaluate to 
\bea\label{untwpartfnfullbar}
\bar Z^{\rm fer, sum}_{(0,0)}(\bar \tau, \bar u, \bar z)   =  \frac{1}{\overline{\eta(\t)^{4}}} \, \overline{ \vartheta_{11}(\tau,  z/2- u)^{2} } \, 
\overline{\vartheta_{11}(\tau, z/2)^{2} }  \, .
\eea
Note that
\be\label{zerosofT4}
Z^{\rm fer,sum}_{(0,0)} \bar Z^{\rm fer,sum}_{(0,0)} \sim z^2 \bar z^2 ~~~~{\rm as} ~ z \rightarrow 0 \, . 
\ee

\paragraph{The twisted sectors}

In the twisted sector $(r,s) \neq (0,0)$, we obtain:
\bea\label{twpartfnrs}
Z_{(r,s)}(\t , \tbar, u, \ubar, z, \zbar) & = & 8 \sqrt{2} \, 
\t_{2}^{1/2} \, e^{{-2 \pi u_{2}^{2} / \tau_{2}-2 \pi z_{2}^{2} / \tau_{2}}} \,
e^{{4 \pi i (u_{1}z_{2}+z_{2}u_{1})/ \tau_{2}}} \, 
 \frac{ |\eta(\t)^{2} |^{2} }{|\vartheta_{1}(\tau,u)|^{2}} \times \, \\
&& \qquad \times \; \bigg| \frac{\eta(\t)^{2}}{\vartheta_{11}(\t,(s+r\t)/2)^{2}} \bigg|^{2}  \, Z^{\rm fer, sum}_{(r,s)}(\tau, u, z) \, 
 \bar Z^{\rm fer, sum}_{(r,s)}(\tbar, \ubar, \zbar) \, . \nonumber
\eea
The left-moving fermion partition functions involve a sum over the various fermion periodicities and 
in each case, a Riemann theta identity (see Appendix \ref{thetaids}) allows us to sum them up into a product 
form. They are given by:
\bea\label{K3partfnpieces}
Z^{\rm fer, sum}_{(0,1)}(\tau, u, z) & = & \frac{1}{\eta(\t)^{4}} \,  \, \vartheta_{11}( \tau, z/2)^{2}  \,   
\vartheta_{10} ( \tau, z/2 +u)^{2} \, , \cr
Z^{\rm fer, sum}_{(1,0)} (\tau, u, z) & = & \frac{1}{\eta(\t)^{4}} \,  \, \vartheta_{11}( \tau, z/2)^{2}  \,   
\vartheta_{01} ( \tau, z/2 + u)^{2} \, , \\
Z^{\rm fer, sum}_{(1,1)} (\tau, u, z) & = & \frac{1}{\eta(\t)^{4}} \, \vartheta_{11}( \tau, z/2)^{2}  \,   
\vartheta_{00} ( \tau, z/2 +u)^{2}  \, . \nonumber
\eea
On the right-moving side, we get:
\bea\label{K3partfnpiecesbar}
\bar Z^{\rm fer, sum}_{(0,1)} (\bar \tau, \bar u, \bar z) & = & \frac{1}{\overline{\eta(\t)^{4}}} \, \overline{\vartheta_{11} ( \tau, z/2-u)^{2}} \,   
\overline{\vartheta_{10} ( \tau, z/2)^{2}} \, , \cr
\bar Z^{\rm fer, sum}_{(1,0)} (\bar \tau, \bar u, \bar z) & = & \frac{1}{\overline{\eta(\t)^{4}}} \, \overline{\vartheta_{11} ( \tau, z/2-u)^{2}} \,   
\overline{\vartheta_{01} ( \tau, z/2)^{2}} \, , \\
\bar Z^{\rm fer, sum}_{(1,1)} (\bar \tau, \bar u, \bar z) & = & \frac{1}{\overline{\eta(\t)^{4}}} \, \overline{\vartheta_{11}( \tau, z/2-u)^{2}} \,   
\overline{\vartheta_{00} ( \tau, z/2)^{2}} \, .\nonumber
\eea
Note that
\be\label{zto0twisted}
Z^{\rm fer,sum}_{(r,s)} \bar Z^{\rm fer,sum}_{(r,s)} \sim z^2 ~~~~{\rm as} ~ z \rightarrow 0 \, . 
\ee

\subsection{Helicity supertrace} \label{helstr}

Now we are in a position to compute the helicity supertraces:
\be
\chi_{n}(\t)  \=  \Tr \, J_{sp}^{n} \, (-1)^{F_{s}} \, q^{L_{0}} \= \int_{E(\t)} \frac{du_{1} du_{2}}{\t_{2}} \;
\big(\frac{1}{2\pi i}(\p_{z} - \p_{\bar z})\big)^{n} \, Z(\t,u, \ubar, z, \zbar) \Big|_{z=\zbar =0} \, ,
\ee
with 
\be
Z(\t,u, \ubar, z, \zbar)  \=   \sum_{r,s=0,1}  Z_{(r,s)}(\t , \tbar, u, \ubar, z, \zbar)  \, .
\ee

From~\eqref{zerosofT4}, it is clear that the untwisted partition function $Z_{(0,0)}$
\eqref{untwpartfn}-\eqref{untwpartfnfullbar} does not contribute to $\chi_{0}$ and $\chi_{2}$, 
and the first non-vanishing result to which it contributes is $\chi_{4}$. 
This is consistent with the fact that the untwisted sector has the same number 
of fermion zero modes as the theory on $T^{4}$.
Similarly, it is clear from~\eqref{zto0twisted} that the twisted sector partition 
functions~$Z_{(r,s)}$ \eqref{twpartfnrs}--\eqref{K3partfnpiecesbar} do not contribute to $\chi_{0}$ 
but they do contribute to~$\chi_{2}$.

The first non-vanishing result is thus $\chi_{2}$, and this receives contributions only from the sectors with $(r,s) \ne (0,0)$:
\bea\label{K3partfn}
\sum_{(r,s)\neq (0,0)}  Z_{(r,s)}(\t , \tbar, u, \ubar, z, \zbar) & = & 8 \sqrt{2} \, 
\t_{2}^{1/2}  \, e^{{-2 \pi u_{2}^{2} / \tau_{2}-2 \pi z_{2}^{2} / \tau_{2}}} \, e^{{4 \pi i (u_{1}z_{2}+u_{2}z_{1})/ \tau_{2}}} \, 
 \frac{1}{|\vartheta_{1}(\tau,u)|^{2}}  \, \times \\
& & \;  \times \; \vth_{1}(\t,z/2)^{2} \, \, \overline{\vth_{1}(\t,z/2-u)}^{2} \sum_{i=2,3,4} \frac{\vth_{i}(\t,z/2+u)^{2}}{\vth_{i}(\t,0)^{2}} \, \frac{\overline{\vth_{i}(\t, z/2)^{2}}}{\vth_{i}(\t,0)^{2}}  \nonumber \, .
\eea

We have:
\bea
&& \big(\frac{1}{2\pi i}(\p_{z} - \p_{\bar z})\big)^{2} \, Z(\t,u, \ubar, z, \zbar) \Big|_{z=\zbar =0}  \=  
 \big(\frac{1}{2\pi i} \p_{z}\big)^{2} \, Z(\t,u, \ubar, z, \zbar) \Big|_{z=\zbar =0} \cr
& & \quad \qquad \=  4 \sqrt{2} \,  \, \t_{2}^{1/2} \, e^{{-2 \pi u_{2}^{2} / \tau_{2}}} \,  
\frac{1}{|\vartheta_{1}(\tau,u)|^{2}} \, \eta(\t)^{6} \, \overline{\vth_{1}(\t, u)^{2}} \, 
\sum_{i=2,3,4} \frac{\vth_{i}(\t,z/2+u)^{2}}{\vth_{i}(\t,0)^{2}}
 \, , \cr
& &  \quad \qquad \= 4 \sqrt{2} \,   \, \t_{2}^{1/2} \, e^{{-2 \pi u_{2}^{2} / \tau_{2}}} \,  
\frac{\eta(\t)^{6}}{\vartheta_{1}(\tau,u)}   \, \overline{\vth_{1}(\t, u)} \, 
\sum_{i=2,3,4} \frac{\vth_{i}(\t,u)^{2}}{\vth_{i}(\t,0)^{2}} \, . 
\eea
Note that although we started with a full string theory with all the fermion periodicities, the spacetime 
computation is such that after summing over all the twisted sectors, the final answer only depends 
on\footnote{A similar phenomenon happens in the computation of helicity supertraces in 
flat space~\cite{Kiritsis}.} the elliptic genus of $K3$ 
\be
{\cal Z}^{\rm ell}(K3;\t,u) \,  \= 8 \sum_{i=2,3,4} \frac{\vth_{i}(\t,u)^{2}}{\vth_{i}(\t,0)^{2}} \, . 
\ee
We thus obtain our main result for the second helicity supertrace:
\be \label{chi2int}
\chi_{2}(\t) \=  \frac{1}{2} \int_{E(\t)} \frac{du_{1} du_{2}}{\t_{2}} \,  (2 \t_{2})^{1/2} \, e^{{-2 \pi u_{2}^{2} / \tau_{2}}} \; 
\frac{\eta(\t)^{6}}{\vartheta_{1}(\tau,u)}   \; \overline{\vth_{1}(\t, u)} \; {\cal Z}^{\rm ell}(K3;\t,u) \, . 
\ee

It is useful to  rewrite the integral \eqref{chi2int} in the language of Jacobi forms
We first write down some notation and standard facts~\cite{Eichler:1985ja} that will be useful. 
A Jacobi form is a holomorphic function $\v(\tau, u)$ from $\mathbb{H} \times\IC$ to $\IC$ which 
is ``modular in $\tau$ and elliptic in $u $'' in the sense that it transforms under the modular group as
  \be\label{modtransform}  \v\Bigl(\frac{a\t+b}{c\t+d},\frac{u}{c\t+d}\Bigr) \= 
   (c\t+d)^k\,e^{\frac{2\pi imc u^2}{c\t+d}}\,\v(\t,u)  \qquad \forall \quad
   \Bigl(\begin{array}{cc} a&b\\ c&d \end{array} \Bigr) \in SL(2; \Z) \ee
and under the translations of $u$ by $\mathbb{Z} \tau + \mathbb{Z}$ as
  \be\label{elliptic}  \v(\t, u+\lambda\tau+\mu)\= e^{-2\pi i m(\lambda^2 \t + 2 \lambda u)} \v(\t, u)
  \qquad \forall \quad \lambda,\,\mu \in \Z \, , \ee
where $k$ is an integer and $m$ is a positive integer. We denote Jacobi forms of weight $k$ and index $m$ by $\v_{k,m}$.
The ring of Jacobi forms of even weight is generated by the two Jacobi forms (our conventions for theta functions are given in Appendix \ref{thetaids}).
\be \label{Jacbasis}
\v_{-2,1}(\t,u) \= \frac{\vth_{1}(\t,u)^{2}}{\eta(\t)^{6}}  \, , \qquad 
\v_{0,1}(\t,u) \= 4 \sum_{i=2,3,4} \frac{\vth_{i}(\t,u)^{2}}{\vth_{i}(\t,0)^{2}} \, .
\ee
The function
\be
P(\t,u) =  \frac{\v_{0,1}(\t,u)}{\v_{-2,1}(\t,u)} = - \frac{3}{\pi^2} \wp(\tau,u)
\ee
with $\wp(\tau,u)$ the usual Weierstrass function is a Jacobi form of weight $2$ and index $0$, which implies that it is invariant under the elliptic transformations \eqref{elliptic} of the Jacobi group. It has double poles of residue $-3/\pi^{2}$ at $z=0$ and its translates by the lattice~$\IZ \t + \IZ$.  We have chosen this normalization to streamline the notation here and in the  manipulations of the integrals in Appendix A.
We also define the non-holomorphic function:
\be
H(\t,u) \= (2\t_{2})^{1/2} e^{-2 \pi u_{2}^{2}/\t_{2}} | \vth_{1}(\t,u) |^{2} \, ,
\ee
which is  invariant under the full Jacobi group as can be easily checked. We then have
\be \label{chi2intPH}
\chi_{2}(\t) \= \int_{E(\t)} \frac{du_{1} du_{2}}{\t_{2}}  \,  P(\t,u) \, H(\t,u) \, .
\ee

This integral has been evaluated by Gaiotto and Zagier \cite{GZ}. We present a brief analysis and a slightly 
different method of evaluation in Appendix~\ref{Appintegral}.  The result is:
\be
\chi_{2}(\t) = - \frac{1}{2} \, \eta(\t)^{3} \, \widehat H^{(2)}(\tau) \, .
\ee
where $\widehat H^{(2)}(\tau)$ is the modular completion of a mock modular form discussed in the introduction. 
We discuss the notion of the modular completion in the following section.

\section{The mock theta function $H^{(2)}(\tau)$ and the twisted BPS index}

In this section we give a quick summary of the definition of  mock modular forms and  of the mock theta function which appeared in the previous section  in the computation
of the BPS index $\chi_2(\tau)$.  We then consider the effects of twisting. 

A holomorphic function $h(\tau)$ on the upper half plane $\HH$ is called a weakly holomorphic mock modular form of weight $k$ for $\Gamma_1=SL_2(\ZZ)$ if it has
at most exponential growth as $\tau \rightarrow i \infty$ and if there exists a modular form $g(\tau)$ of weight $k-2$ on $\Gamma_1$ such that the \emph{completion} of
$h(\tau)$ given by
\be \label{shad}
\widehat h(\tau) = h(\tau) +(4i)^{k-1} \int_{- \bar \tau}^\infty (z+ \tau)^{-k} \overline{g(-\bar z)} dz
\ee
transforms like a holomorphic modular form of weight $k$ on $\Gamma_1$ with some multiplier system~$\nu$. The modular form $g(\tau)$ is called the {\it shadow} of the
mock modular form $h(\tau)$. The completion~$\wh h$ obeys 
    \be\label{ddtbarh}   4 i \, (\t_2/2)^k\,\,\frac{\pa \wh h(\t)}{\pa \bar{\tau}} 
    \= \bar{g(\tau)}\;.  \ee
When the shadow~$g$  is a unary theta series of weight 1/2 or 3/2, then the mock modular form~$h$ is called 
a mock theta function of weight 3/2 or 1/2, respectively. 

The example appearing in this paper is the mock theta function $H^{(2)}(\tau)$ which appeared in the physics literature in the decomposition of the elliptic genus of
$K3$ in terms of characters of the $N=4$ superconformal algebra \cite{eot}. It can also be defined as follows~\cite{DMZ}. Let
\be
F_2^{(2)}(\t) \=  \sum_{r>s>0 \atop r-s \, \rm odd} (-1)^r\,s\,q^{rs/2} \= q + q^2 -  q^3 +  q^4 - q^5 + \cdots \, .
\ee
Then the function $H^{(2)}$ and its Fourier coefficients $c^{(2)}$ are defined by:
\begin{align}
H^{(2)}(\tau) &= \frac{ 48 F_2^{(2)}(\tau) - 2 E_2(\tau)}{ \eta(\tau)^3}= \sum_{n=0}^\infty c^{(2)}(8n-1) \, q^{n-1/8} \\
&= 2q^{-1/8} \biggl(-1 + 45 q + 231 q^2+ 770 q^3 + 2277 q^4 + \cdots \biggr)
\end{align}
where $E_2(\tau)$ is the usual Eisenstein series and $\eta(\tau) = q^{1/24} \prod_{n=1}^\infty(1-q^n)$ is the Dedekind  eta function. $H^{(2)}(\tau)$ is a weight $1/2$ mock
modular form with shadow $24 \, \eta(\tau)^3$ and a multiplier system conjugate to that of $\eta(\tau)^3$. Evaluating the integral in \eqref{shad} gives an explicit formula for
the completion
\be
\widehat H^{(2)}(\tau) = H^{(2)}(\tau) + 24 \sum_{k \in \ZZ} {\rm sgn}(4k+1) q^{-(4k+1)^2/8} \biggl( -1 + {\rm Erf} \biggl[ \frac{4k+1}{2} \sqrt{2 \pi \tau_2} \biggr]  \biggr)
\ee
with $\tau_2$ the imaginary part of $\tau$ and ${\rm Erf}[x]$ the error function. 

The first few coefficients $c^{(2)}(8n-1)$ in the $q$-expansion of $H^{(2)}(\tau)$ are dimensions of irreducible representations of the Mathieu group $M_{24}$ \cite{eot}.
It is natural to think that each coefficient $c^{(2)}(8n-1)$ should be identified with the dimension of an $M_{24}$ module $K_n$ so that $c^{(2)}(8n-1)={\rm dim}K_n= {\rm Tr}_{K_n} 1$. This idea by itself is ambiguous because there are many possible decompositions of the coefficients into dimensions of irreducible representations (irreps) of $M_{24}$.
To test the idea one follows the same logic as in the computation of the McKay-Thompson series of Monstrous Moonshine \cite{Thompson, conway_norton} and studies
the series $H^{(2)}_g(\tau) = \sum_{n} {\rm Tr}_{K_n} g \, q^{n-1/8}$ for $g \in M_{24}$. These McKay-Thompson series depend only on the conjugacy class of $g$, and if for each
conjugacy class the $H^{(2)}_g(\tau)$ are also mock modular forms this is interpreted as positive evidence for a correct choice of decomposition into irreps as well as for a moonshine connection between the mock modular form $H^{(2)}(\tau)$ and the finite simple group $M_{24}$.  This strategy has been used  in \cite{cheng,GaberdielI,GaberdielII,EguchiI} to compute the mock modular forms
$H^{(2)}_g(\tau)$ for all conjugacy classes of $M_{24}$ and thus  determine the decomposition of the coefficients $c^{(2)}(8n-1)$ into irreps of $M_{24}$ .  Using the notation of the review \cite{cd_rev} the resulting mock modular forms can be written in the form
\be \label{twist_h}
H^{(2)}_g(\tau) = \frac{\chi(g)}{24} H^{(2)}(\tau) - \frac{\widetilde T_g(\tau)}{\eta(\tau)^3}
\ee
where $\chi(g)$ is the character of $g$ in the $24$-dimensional permutation representation of $M_{24}$ with a decomposition $24 = 23 \oplus 1$ in terms of irreps. 
Here the $\widetilde T_g(\tau)$ are a set of weight two modular forms for congruence subgroups which can be found tabulated in~\cite{cd_rev} and the $H_g^{(2)}(\tau)$ are weight $1/2$
mock modular forms  for $\Gamma_0(N_g)$ with shadow $\chi(g) \,\eta(\tau)^3$. The number~$N_g$ is an integer known as the level of $g$ and
determined by the cycle shape of $g$ in the 24-dimensional permutation representation of $M_{24}$. See the review  \cite{cd_rev} for details.

At special points in the moduli space of~$K3$, one has a SCFT description of the $K3$ surface. 
At such points, all the discrete symmetries of~$K3$ that preserve supersymmetry can be classified \cite{Gaberdiel:2011fg}. 
This list includes and extends the symplectic automorphisms of the~$K3$ 
surface that were classified by Mukai \cite{Mukai} and by Kondo \cite{Kondo}, but 
does not include all elements of~$M_{24}$.
For elements $g \in M_{24}$ that  are within this class, one has a somewhat better understanding of the 
McKay-Thompson series~\eqref{twist_h}. 
Using the SCFT description, one can compute a twisted version of the elliptic genus:
\be \label{twell}
{\cal Z}^{\rm ell}_g(K3;\tau,u) \= \Tr_{RR} \, (-1)^{F} \, g \, q^{L_{0}-c/24} \, \bar q^{\wt L_{0}-\wt c/24} \zeta^{J_{0}} \, , 
\ee
where the trace is over the RR sector of the Hilbert space of the $K3$ SCFT. 
The twisted elliptic genus ${\cal Z}^{\rm ell}_g(K3;\tau,u) $ is also a Jacobi form over a subgroup of the 
full Jacobi group. Using this fact, one can decompose it into the basis elements~\eqref{Jacbasis} with coefficients
being modular forms on a subgroup of~$SL(2,\IZ)$. For all elements~$g$ for which the twisted elliptic genus 
has been computed, one finds \cite{cd_rev}:
\be \label{twist_ell}
{\cal Z}^{\rm ell}_g(K3;\tau,u) = \frac{\chi(g)}{24} {\cal Z}^{\rm ell}(K3;\tau,u) + \widetilde T_g(\tau) \varphi_{-2,1}(\tau,u) \, . 
\ee
On decomposing the twisted elliptic genus into characters of the $N=4$ superconformal algebra and throwing out 
the massless representation as before, one obtains the McKay-Thompson series~$H^{(2)}_g(\tau)$.

From the point of view of this paper, the NS5-brane system naturally produces the McKay-Thompson 
series~$H^{(2)}_g(\tau)$. The integral in Eqn.~\eqref{chi2int} that gives us a map from the elliptic genus of $K3$ to the completion of the  weight two mixed mock modular form 
$-(1/2) \eta(\tau)^3 H^{(2)}(\tau)$ that
can be obviously generalized to a map from the twisted form of the elliptic genus given in Eqn.~\eqref{twist_ell} to a twisted version of the completion.  We can check that this correctly leads to the twisted mock
modular form $H^{(2)}_g(\tau)$ as follows. We define
\be \label{chigint}
\chi_{2,g}(\tau) = \int_{E(\tau)} P_g(\tau,u) H(\tau,u) \frac{du_1 du_2}{\tau_2}
\ee
with
\be
P_g(\tau,u) = \frac{1}{2} \frac{{\cal Z}^{\rm ell}_{g}(K3;\tau,u)}{\varphi_{-2,1}(\tau,u)} \, .
\ee
Then using Eqn.~\eqref{twist_ell} and Eqn.~\eqref{twist_h}  as well as the integral in Eqn.~\eqref{hint} of Appendix A we find
\be
\chi_{2,g} = - \frac{1}{2} \eta(\tau)^3 \widehat H^{(2)}_g(\tau)
\ee
where $\widehat H^{(2)}_g(\tau)$ is the completion of $H^{(2)}_g(\tau)$.  Thus the map from twisted elliptic genera to twisted mock modular forms provided by the
integral in Eqn.~\eqref{chi2int} agrees with the map given by the decomposition of the twisted elliptic genus into characters of the $N=4$ superconformal algebra.

Further, the superstring computation in \S\ref{compBPS} that led to the integral in Eqn.~\eqref{chi2int} can itself 
be generalised to include the twist~$g$. If~$g$ is a symmetry of the~$K3$ SCFT that preserves the worldsheet 
supersymmetry, then it can be lifted to a corresponding symmetry of the superstring theory 
discussed in \S\ref{superstring}, and we 
can compute 
\be \label{defchig2}
\chi_{2,g}(\t) \= \Tr \, (J_{\rm sp})^{2} \, (-1)^{F_{\rm s}} \, g \, q^{L_{0}- c/24} \, \bar q^{\wt L_{0}-\wt c/24}  
\ee
in this superstring theory. The sum over NS and R sectors with the insertion of the GSO projection for the 
twisted superstring index collapses as before in such a way that the final answer only depends on the twisted 
SCFT elliptic genus~\eqref{twell}. The main technical point here is that the sum over NS and R sectors with the 
GSO projection involves eight free worldsheet fermions, and the Riemann identities used in~\S\ref{helstr} to sum 
the various expressions are an manifestation of spacetime supersymmetry, as is the case for superstring theory 
in 10 flat dimensions. We can identify the spinorial charges that the spacetime supercharges have under the 
various rotational symmetries of the theory, but we have not explicitly constructed the Green-Schwarz superstring 
for the cigar theory (see~\cite{Murthy:2003es} for some more discussion of this subject).

More generally, since we are working at the level of superconformal field theory, we can consider automorphisms of the full superconformal field theory~\eqref{wsSCFT} which preserve spacetime supersymmetry. These transformations certainly include such symmetries of the $K3$ component  of our superconformal field theory as were analyzed in \cite{Gaberdiel:2011fg}. The full extension of this classification to the superconformal field theories  considered here
is a very interesting problem that we hope to return to in the future. 

\section{Discussion and conclusions \label{discussion}}

As mentioned in the introduction, our goal in this paper was to find a BPS state counting problem in string theory that leads to the mock
modular form $H^{(2)}(\tau)$ (or its modular completion) and we suggested that the required construction would remove the massless string states from the spectrum.
The two NS5-brane system on $K3 \times S^1$ achieves what
we want in a natural manner, but the connection to our earlier discussion may not be completely clear so here we make some further remarks on
out interpretation of the calculation performed in this paper.

The K3 elliptic genus can be written in terms of Jacobi forms as (see e.g \cite{eguchihikami} or Eqn.~(7.39) of~\cite{DMZ} )
\be \label{elldecomp}
{\cal Z}^{ell}(K3, \tau,u) = 2 \varphi_{0,1}(\tau,u) = -24 \mu(\tau,u) \eta(\tau)^3  \varphi_{-2,1}(\tau,u) - \eta(\tau)^3 H^{(2)}(\tau) \varphi_{-2,1}(\tau,u)
\ee
with
\be
\mu(\tau,u) =\frac{e^{\pi i u}}{\vartheta_1(\tau,u)} \sum_{n \in \ZZ} \frac{(-1)^n q^{(n^2+n)/2} e^{2 \pi i n u}}{1-q^n e^{2 \pi i u}} \, .
\ee
The first term on the right hand side of \eqref{elldecomp} is related to  a massless character of the world sheet $N=4$ superconformal algebra and from a spacetime point of view encodes  the massless graviton degree of freedom and its descendants. These modes have wave functions that are delocalized along the length of the cigar. In comparison, the second term corresponds to massive modes that are localized near the tip of the cigar. The holomorphic mock modular form~$H^{(2)}$ counts the localized modes (up to a factor of
$- (1/2) \eta(\tau)^3)$, while the delocalized modes contribute to the  non-holomorphic part of  the full  BPS index ~$-(1/2) \eta(\tau)^3  \wh H^{(2)}$. The 5-brane background and the process of taking the near horizon limit
has in a sense removed some of the massless modes associated to the first term in~\eqref{elldecomp} which give a holomorphic term such that the sum in \eqref{elldecomp} is
modular, and replaced them by a set of delocalized modes which give a non-holomorphic contribution which also leads to a modular answer. 

Naively we would expect the BPS state counting formula to be holomorphic based 
on the argument of pairing of bosonic and fermionic modes while the answer we obtain is clearly  not holomorphic. The resolution of this puzzle arises from recent studies of non-compact SCFTs in which such a phenomenon has been 
unravelled~\cite{TroostI,ESII,Ashok}. The point is that the non-compactness requires us to specify normalizability conditions for all the modes in the spectrum, and supersymmetry does not commute with these conditions. Note that the form of the spacetime supercharges that we write down in Eqn.~\eqref{susyalg} are only valid in the asymptotic region of the cigar, and their exact form is more complicated. From a technical point of view, the non-compactness produces a continuum and an associated density of states of bosons and fermions that are not equal. The difference in the density of states is proportional to the reflection coefficient of a wave sent down the throat of the cigar~\cite{TroostI}. 

%NEW here
A notion of holography exists for the theory of NS fivebranes in string theory~\cite{Aharony:1998ub, Giveon:1999zm}. From this 
point of view, we expect that the BPS states studied in this paper are related to the BPS states of the 
non-gravitational low-energy theory of the fluctuations of the fivebranes wrapped on~$K3$. 
Theories of fivebranes in M-theory wrapping various two and four dimensional surfaces have generated great interest in 
the last few years following the work of~\cite{Gaiotto} and it would be very interesting to make this relation precise.

If we had not taken the near-horizon limit of the NS5-brane, but instead looked for  bound states of NS5-branes with  fundamental strings carrying momentum, we would have
obtained a BPS three charge black hole with a macroscopic horizon size in five dimensional asymptotically flat space. It would be very interesting to understand the relation of these ``big'' black holes to the counting problem we have analyzed and thus possibly to moonshine. 
Three  charge BPS black holes in five dimensions are also closely related to four-charge black holes in four dimensions that exhibit  the wall-crossing phenomenon. Mathematically  they are described by a family of mock modular forms~\cite{DMZ} that are a priori unrelated to the mock modular form that we study in this paper. It would be interesting to find relations between the mock modular forms appearing in these two counting problems. 

Finally, there are several obvious generalizations of the present work that we hope to return to in the near future. One of these is the extension of our analysis to an arbitrary number of fivebranes. It would be particularly interesting to see if there is any connection between the ADE classification of fivebranes and the ADE classification which appears in the analysis of umbral moonshine \cite{mum}.  Another promising direction involves the  computation of the BPS index for CHL models constructed as $(K3 \times S^1)/(\ZZ/n)$ where $\ZZ/n$ acts as an order $n$ shift on the $S^1$ and as an order
$n$ symplectic automorphism of $K3$. Finally, it would interesting to analyze  the full group of supersymmetry preserving automorphisms for the BPS configuration analyzed here 
and its generalization to CHL models and arbitrary numbers of fivebranes.

\acknowledgments

We thank Don Zagier for sharing with us the results of his unpublished work with D.~Gaiotto.  JH acknowledges the support of NSF grant 1214409 and the hospitality of the theory group at Nikhef and the Aspen Center for Physics during portions of this work.
The work of SM is supported by the ERC Advanced Grant no. 246974,
{\it ``Supersymmetry: a window to non-perturbative physics''}.

\appendix

\section{Analysis of the integral for the second helicity supertrace $\chi_{2}$} \label{Appintegral}

In this appendix, we analyze and evaluate the integral \eqref{chi2intPH} that gives the second helicity supertrace. 
In terms of the functions 
\be
P(\t,u) =  \frac{\v_{0,1}(\t,u)}{\v_{-2,1}(\t,u)} = - \frac{3}{\pi^2} \wp(\tau,u) 
\ee
and 
\be
H(\t,u) \= (2\t_{2})^{1/2} e^{-2 \pi u_{2}^{2}/\t_{2}} | \vth_{1}(\t,u) |^{2} \, ,
\ee
the integral is written as:
\be \label{chi2Jac}
\chi_{2}(\t) \= \int_{E(\t)}   P(\t,u) \, H(\t,u) \, \frac{du_{1} du_{2}}{\t_{2}}  \, .
\ee
where $E(\tau)$ is the elliptic curve $\CC/(\ZZ \tau + \ZZ)$.  We use the notation $q=e^{2 \pi i \tau}$, $y=e^{2 \pi i u}$.

On the right-hand side of this equation, the integrand, the integration region, 
and the measure are all manifestly invariant under the elliptic transformations. Further, the integrand is a 
(meromorphic) Jacobi form of weight 2. If the integral is well-defined, it is thus manifest that the 
function~$\chi_{2}(\t)$ transforms as a holomorphic modular form of weight~$k=2$. We say ``transforms as'',
and not ``is'' a holomorphic modular form because, as we shall see below, the function $\chi_{2}$ is not 
holomorphic in~$\t$, it is the non-holomorphic completion of a (mixed) mock modular form.

We now show that the integral \eqref{chi2Jac} is well defined. 
The only possible problems come from the behavior as $u$ approaches $0,1, \tau, \tau+1$. To analyze the behavior near these points we cut
out a pizza slice of radius~$\varepsilon <<1$ around each of these points in $E(\tau)$ so that $E(\tau) = 
E^{\varepsilon} + D^{\varepsilon}$ and then study the limit $\varepsilon \rightarrow 0$. Here $E^{\varepsilon}$ is the
``ticket-shaped" region obtained by removing the pizza slices from $E(\tau)$ and $D^{\varepsilon}$ is the disc of radius $\varepsilon$ formed by assembling
the four slices into a single disc of radius $\varepsilon$ at the origin using the elliptic invariance of the integrand.

Now consider the integral over the disc $D^{\varepsilon}$. As $u \rightarrow 0$ we have $P(\t,u) \sim u^{-2} $ and
$\vartheta_1(\tau,u) \sim u$. Therefore the only potentially problematic part of the integrand is
\be
\int_{D^{\varepsilon}} e^{-2 \pi u_2^2/\tau_2} \, \frac{\bar u}{u} \, du_{1} du_{2} \,.
\ee
Using polar coordinates $u=\rho e^{i \theta}$ this becomes
\begin{align}
& \int_0^{2 \pi} d \theta \int_0^\varepsilon \rho \, d \rho \, e^{-2 \pi \rho^2 \sin^2 \theta/\tau_2} e^{-2 i \theta} \\
& =-\frac{\tau_2}{4 \pi}  \int_0^{2 \pi} d \theta \, \frac{e^{-2 i \theta}}{\sin^2\theta} \, (1- e^{- 2 \pi \varepsilon^2 \sin^2 \theta/\tau_2}) = \frac{\pi^2}{4 \tau_2} \varepsilon^4 + O(\varepsilon^6)  \, .
\end{align}
Since this vanishes as $\varepsilon \rightarrow 0$ we can simply define the integral as
\be \label{chi2def}
\chi_{2} (\t) \=   \lim_{\varepsilon \rightarrow 0}  \int_{E^{\varepsilon}} H(\tau,u) \, P(\tau,u) \, \frac{du_{1} du_{2}
}{\tau_2} \, .
\ee
Since $P(\tau,u)$ is analytic in the region $E^{\varepsilon}$ we can safely set $\partial_{\bar u} \, P(\tau,u)=0$ inside the integral in the manipulations below. 

We now compute the $\tbar$ derivative of the function~$\chi_{2}$. By a change of variables~$u=a \tau + b$, we 
have:
\be
\chi_2(\t) \= \int_{0}^{1} \int_{0}^{1} H(\t,a\t +b) \, P(\t,a\t +b) \, db \, da \, . 
\ee
Since $P$ is meromorphic in $\t$, the only local $\tbar$ dependence comes from the function $H$.
We have:
\be\label{Ftbar}
\p_{\tbar} \, \chi_2(\t)  \= \int_{0}^{1} \int_{0}^{1} (\p_{\tbar} H(\t,a\t +b)) \, P(\t,a\t +b) \, db \, da \, . 
\ee
One can check that:
\be \label{pbarH}
\p_{\tbar} H(\t,a,b)\, \equiv \, \p_{\tbar} H(\t,a\t+b) \= \Big( \frac{i}{4 \pi} \p_{\bar u}^{2} H(\t,u) \Big)_{u=a\t+b} \, . 
\ee
Plugging \eqref{pbarH} into \eqref{Ftbar}, and changing variables to $u = u_{1} + i u_{2},\bar u = u_{1}-iu_{2}$, we obtain:
\bea \label{Ftbar2}
\p_{\tbar}\chi_2(\t)  &\=& \frac{1}{8 \pi} \int_{E^{\varepsilon}} (\p_{\bar u}^{2} H(\t,u)) \, P(\t,u) \, \frac{d\bar u \, du}{\t_{2}}  \cr
& \= &  \frac{1}{8 \pi} \int_{E^{\varepsilon}}  \p_{\bar u} \Big(  \p_{\bar u} H(\t,u) \, P(\t,u) \, \frac{1}{\t_{2}}  \Big) \, d\bar u \, du \cr
& \= & \frac{1}{8 \pi} \oint_{\p E^{\ve}}  \p_{\bar u} H(\t,u) \, P(\t,u)   \, \frac{1}{\t_{2}} \, du \, . 
\eea
The integral along the four straight edges of $E^{\ve}$ adds up to zero since we go around the opposite 
straight edges in opposite directions, and the integrand is equal by elliptic invariance. Therefore we have:
\bea \label{Ftbar3}
\p_{\tbar}\chi_2(\t)  &\=& -\frac{1}{8 \pi} \oint_{\p D^{\ve}}  \p_{\bar u} H(\t,u) \, P(\t,u)   \, \frac{du}{\t_{2}} \, \cr
 & = &  -\sqrt{2} \frac{1}{8 \pi \tau_2} (2 \pi i) \, {\rm Res}_{u \rightarrow 0} \biggl( \partial_{\bar u} H(\tau,u) \frac{1}{u^2} \biggr) \cr
 & = & -i \sqrt{2} \frac{1}{4 \tau_2} \biggl( \partial_u \partial_{\bar u} H(\tau,u) \biggr)_{u=0} = i \sqrt{2} \, \pi^2 \tau_2^{-1/2} \eta(\tau)^3 \, \overline{\eta(\tau)^3} \, .
\eea

The function $\chi_{2}/\eta^{3}$ transforms as a holomorphic modular form of weight~$k=1/2$ 
and the above shows that it obeys the holomorphic anomaly equation:
\be
\frac{1}{i \sqrt{2}\pi^2 } \, \tau_2^{1/2} \, \p_{\tbar} \frac{\chi_2(\t)}{\eta(\t)^{3}}  \= \overline{\eta(\tau)^3} \, .
\ee
In other words, $\chi_{2}/\eta^{3}$ is a mock modular form of weight~$k=1/2$ and shadow~$-12 \eta^{3}$.

Following \cite{GZ}  we now evaluate this integral and find 
\be
\chi_{2}(\t) = - \frac{1}{2} \, \eta(\t)^{3} \, \widehat H^{(2)}(\tau) \, .
\ee
As a first step towards this result we show that
\be \label{hint}
I^{(2)}(\tau)= \int_{E(\tau)} H(\tau,u) \frac{du_1 du_2}{\tau_2}=1 \, .
\ee
We use the expansion
\be
|\vartheta_1(\tau,u)|^2 = \sum_{n,m \in \ZZ} q^{(n+1/2)^2/2} {\bar q}^{(m+1/2)^2/2} e^{2 \pi i[(n+1/2)(u+1/2) - (m+1/2)(\bar u+1/2)]}
\ee
and write $\tau,u$ in terms of real and imaginary parts $\tau=\tau_1 + i \tau_2$, $u=u_1+i u_2$
to give
\begin{align}
|\vartheta_1(\tau,u)|^2 & = \sum_{n,m  \in \ZZ} {\rm Exp} \biggl[ 2 \pi i \biggl( \frac{\tau_1}{2}(n^2-m^2+n-m)+i \frac{\tau_2}{2}(n^2+m^2+n+m+\frac{1}{2})  \\
& + u_1(n-m)+i u_2(n+m+1)+\frac{n-m}{2} \biggr) \biggr] \, .
\end{align}
Now change variables from $(u_1,u_2)$ to $(a,b)$ with $u=a \tau+b$. The Jacobian gives a factor of $\tau_2$
and the only term involving $b$ is
\be
\int_0^1 db~ e^{2 \pi i b(n-m)}= \delta_{n,m}
\ee
so we are left with the integral
\begin{align}
I^{(2)}(\tau) &= \sqrt{2 \tau_2}\int_0^1 da \sum_{n \in \ZZ} {\rm Exp}[- 2\pi a^2 \tau_2 - \pi \tau_2(2n^2+2n+1/2) - 2 \pi a \tau_2(2n+1)] \\
             &= \sqrt{2 \tau_2} \int_0^1 da \sum_{n \in \ZZ} {\rm Exp}[-2 \pi \tau_2(a+n+1/2)^2] \\
             &= \sqrt{2 \tau_2}\sum_{n \in \ZZ} \int_n^{n+1} da_n ~{\rm Exp}[- 2 \pi \tau_2(a_n+1/2)^2]  \\
             &=   \sqrt{2 \tau_2} \int_{-\infty}^{+\infty} dx~ {\rm Exp}[- 2 \pi \tau_2(x+1/2)^2]  =1
\end{align}
where we changed variables to $a_n=a+n$ to convert the integral of the sum to a sum of integrals over the interval $[n,n+1]$.

We now move on to the evaluation of the integral \eqref{chi2def}. We first use the identity (see for example \cite{eguchihikami} or  Eqn. 7.39 of \cite{DMZ} )
\be
P(\tau,u)= \frac{ \varphi_{0,1}(\tau,u)}{\varphi_{-2,1}(\tau,u)} = -12 \mu(\tau,u) \eta^3(\tau) - \frac{1}{2}\eta^3(\tau) H^{(2)}(\tau)
\ee
with
\be
\mu(\tau,u) =\frac{e^{\pi i u}}{\vartheta_1(\tau,u)} \sum_{n \in \ZZ} \frac{(-1)^n q^{(n^2+n)/2} e^{2 \pi i n u}}{1-q^n e^{2 \pi i u}} \, .
\ee
Substituting this into the integrand gives
\be
\chi_2(\tau) = -12 \eta^3(\tau)  \int_{E(\tau)} H(\tau,u) \mu(\tau,u) \frac{du_1 du_2}{\tau_2} - \frac{1}{2} \eta^3(\tau) H^{(2)}(\tau) \int_{T(\tau)} H(\tau,u) \frac{du_1 du_2}{\tau_2} 
\ee
which using the earlier result for $I^{(2)}$ gives
\be \label{t2expr}
\chi_2(\tau) = - \frac{1}{2}\eta^3(\tau) \biggl( H^{(2)}(\tau) +24 \int_{E(\tau)} H(\tau,u) \mu(\tau,u) \frac{du_1 du_2}{\tau_2} \biggr) \, . 
\ee

To evaluate the remaining integral first note that the factors of $\vartheta_1$ cancel out so that
\be \label{hmu}
H(\tau,u) \mu(\tau,u)  =(2 \tau_2)^{1/2} e^{-2 \pi u_2^2/\tau_2} \overline{\vartheta_1(\tau,u)}  \sum_{n \in \ZZ} \frac{(-1)^n q^{(n^2+n)/2} e^{2 \pi i (n+1/2) u}}{1-q^n e^{2 \pi i u}} \, . 
\ee

Now $|q^ny|=e^{-2 \pi n \tau_2}e^{-2 \pi u_2}$. Using modular invariance we can choose $\tau$ to be in the usual fundamental domain of
$SL(2,\ZZ)$ so that $\tau_2 \ge \sqrt{3}/2$ and since $u \in E(\tau)$ we have $0 \le u_2 \le \tau_2$. Now we have a wall-crossing type phenomenon.
For $n \ge 0 $ we expand\footnote{For $n \ge 1$ and $ n \le -2$ the  expansions  in Eqn.~\eqref{eexpone} and Eqn.~\eqref{eexptwo} are correct since $|q^n y|<1$ for all $q,y$. They can be extended to
$n \ge 0$ and $n \le -1$ because the contribution from the boundary where $|q^ny|=1$ vanishes due to the prefactor  in  Eqn.~\eqref{hmu}.}
\be \label{eexpone}
\frac{1}{1-q^n y} = \sum_{k=0}^\infty q^{nk} y^k \, , 
\ee
while for $n \le -1$  we can write
\be \label{eexptwo}
\frac{1}{1-q^n y}= - \frac{q^{-n}y^{-1}}{1-q^{-n}y^{-1}} = - \sum_{k=0}^\infty q^{-n(k+1)} y^{-(k+1)} \, . 
\ee

This then gives us
\be
H(\tau,u) \mu(\tau,u)  = (2 \tau_2)^{1/2} e^{-2 \pi u_2^2/\tau_2} \overline{\vartheta_1(\tau,u) } \biggl({\cal S}_<+ {\cal S}_> \biggr) 
\ee
where
\begin{align}
{\cal S}_< &=  - \sum_{n=-\infty}^{-1} \sum_{k=0}^\infty  (-1)^n q^{\frac{n^2-n-2nk}{2}} y^{n-k-1/2}  \, ,  \\
{\cal S}_> &= \sum_{n=0}^\infty \sum_{k=0}^\infty (-1)^n q^{\frac{n^2+n+2nk}{2}} y^{n+k+1/2}\, . 
\end{align}
Writing $\bar \vartheta_1$ as the sum
\be
\overline{\vartheta_1(\tau,u)} = \sum_{m \in \ZZ} {\bar q}^{\frac{(m+1/2)^2}{2}} {\bar y}^{m+1/2} (-1)^m
\ee
gives us an expression for the integral of $ H \mu$ which has two terms:
\begin{align}
I_< & = - \int_{E(\tau)} \frac{du_1 du_2}{\tau_2} \biggl( (2 \tau_2)^{1/2} e^{- 2 \pi u_2^2/\tau_2} \sum_{m \in \ZZ} \sum_{n=-\infty}^{-1} \sum_{k=0}^\infty (-1)^{n+m} q^{\frac{n^2-n-2nk}{2}} {\bar q}^{\frac{m^2+m+1/4}{2}} y^{n-k-1/2} {\bar y}^{m+1/2} \biggr) \, , \\
I_> &= \int_{E(\tau)} \frac{du_1 du_2}{\tau_2} \biggl( (2 \tau_2)^{1/2} e^{- 2 \pi u_2^2/\tau_2} \sum_{m \in \ZZ} \sum_{n=0}^{+\infty} \sum_{k=0}^\infty (-1)^{n+m} q^{\frac{n^2+n+2nk}{2}} {\bar q}^{\frac{m^2+m+1/4}{2}} y^{n+k+1/2} {\bar y}^{m+1/2} \biggr) \, . 
\end{align}

Let's consider $I_>$ first. Changing variables via $u=a \tau+b$ the integral over $b$ gives $\delta_{n+k,m}$ and we are left after some simplifications with
\be
I_> = (2 \tau_2)^{1/2} \sum_{n=0}^\infty \sum_{k=0}^\infty (-1)^k (q \bar q)^{\frac{n^2+n+2nk}{2}} {\bar q}^{\frac{(k+1/2)^2}{2}} \int_0^1 da e^{- 2 \pi a^2 \tau_2} e^{-4 \pi a \tau_2(n+k+1/2)} \, . 
\ee
Similarly we find
\be
I_< = (2 \tau_2)^{1/2} \sum_{n= - \infty}^{-1} \sum_{k=0}^\infty  (-1)^k (q \bar q)^{\frac{n^2-n-2nk}{2}} {\bar q}^{\frac{(k+1/2)^2}{2}} \int_0^1 e^{- 2 \pi a^2 \tau_2} e^{-4 \pi a \tau_2(n-k-1/2)} \, . 
\ee

We can rewrite these expressions in the form
\begin{align}
I_> & = (2 \tau_2)^{1/2} \sum_{k=0}^\infty (-1)^k  {q}^{-\frac{(k+1/2)^2}{2}}   \sum_{n=0}^\infty \int_0^1 da e^{-2 \pi  \tau_2(a+n+k+1/2)^2} \, , \\
I_< & =( 2 \tau_2)^{1/2}  \sum_{k=0}^\infty  (-1)^k {q}^{-\frac{(k+1/2)^2}{2}} \sum_{n= - \infty}^{-1} \int_0^1 da  e^{-2 \pi  \tau_2(a+n-k-1/2)^2}\, . 
\end{align}

Changing variables as before leads to
\begin{align}
I_<+I_> &= (2 \tau_2)^{1/2} \sum_{k=0}^\infty (-1)^k q^{-\frac{(2k+1)^2}{8}} \biggl( \int_{-\infty}^{+ \infty} e^{- 2 \pi \tau_2 x^2} - \int_{-k-1/2}^{k+1/2} e^{-2 \pi \tau_2 x^2} dx \biggr) \\
& =  \sum_{k=0}^\infty (-1)^k q^{-\frac{(2k+1)^2}{8}} \biggl(1 - {\rm Erf}\biggl[\frac{1+2k}{2} \sqrt{2 \pi \tau_2} \biggr] \biggr) \, .
\end{align}

So finally we find after substitution into Eqn.~\eqref{t2expr}
\begin{align}
\chi_2(\tau) &= - \frac{1}{2} \eta^3(\tau) \biggl( H^{(2)}(\tau) +24 \sum_{k=0}^\infty (-1)^k q^{-\frac{(2k+1)^2}{8}} \biggl(1 - {\rm Erf}\biggl[ \frac{1+2k}{2} \sqrt{2 \pi \tau_2} \biggr] \biggr)  \biggr) \\
& = - \frac{1}{2}  \eta^3(\tau) \widehat H^{(2)}(\tau) \, .
\end{align}

\section{Theta function conventions and Riemann identities \label{thetaids}}
The classical Jacobi theta functions are (with $q=e^{2 \pi i \tau}$, $\zeta=e^{2 \pi i z}$)
\begin{align} \nonumber
 \vartheta_{00}(\tau,z) &=\vartheta_3(\tau,z) = \prod_{n=1}^\infty (1-q^n) (1+\zeta q^{n-1/2})(1+\zeta^{-1} q^{n-1/2}) \\ &= \sum_{m \in \ZZ} q^{m^2/2} \, \zeta^m  \, , \\ \nonumber
 \vartheta_{01}(\tau,z) & =  \vartheta_4(\tau,z) = \prod_{n=1}^\infty (1-q^n) (1-\zeta q^{n-1/2})(1-\zeta^{-1} q^{n-1/2}) \\ &= \sum_{m \in \ZZ} e^{\pi i m}  \, q^{m^2/2} \, \zeta^m  \, , \\ \nonumber
 \vartheta_{10}(\tau,z)  &=   \vartheta_2(\tau,z) =q^{1/8} \zeta^{1/2} \prod_{n=1}^\infty (1-q^n)(1+\zeta q^n)(1+\zeta^{-1} q^{n-1}) \\ &  = \sum_{m \in \ZZ} q^{(m+1/2)^2/2} \, 
\zeta^{m+\half} \, , \\ \nonumber
  \vartheta_{11}(\tau,z) & =-i \vartheta_1(\tau,z) = -q^{1/8} \zeta^{1/2} \prod_{n=1}^\infty (1-q^n) (1-\zeta q^n) (1-\zeta^{-1} q^{n-1}) \\ & =  \sum_{m \in \ZZ} e^{\pi i (m+\half)}  \, q^{(m+1/2)^2/2} \, 
\zeta^{m+\half} \, . 
\end{align}

The conventions for $\vartheta_{00},\vartheta_{01},\vartheta_{10}, \vartheta_{11}$ agree with \cite{Mumford} and the conventions for
$\vartheta_i$, $i=1,2,3,4$ agree with the appendix of \cite{um}
Also the above convention for $\vartheta_{11}$ differs from~\cite{Polchinski}. 

Write  $\vartheta_{ab}(x) \equiv \vartheta_{ab}(\tau,x)$
and let
\begin{align}
 & x_1 = \frac{1}{2}(x+y+u+v)  \, ,  \qquad \qquad y_1 = \frac{1}{2}(x+y-u-v) \, , \cr
 & u_1 = \frac{1}{2}(x-y+u-v)  \, , \qquad \qquad v_1 = \frac{1}{2}(x-y-u+v) .
\end{align}
Then we have the following Riemann theta relations, taken from \cite{Mumford}.
\begin{align} 
\nonumber (R5):  \vartheta_{00} \vartheta_{00} \vartheta_{00} \vartheta_{00}- \vartheta_{01} \vartheta_{01} \vartheta_{01} \vartheta_{01}- \vartheta_{10}\vartheta_{10} 
\vartheta_{10}\vartheta_{10}  + \vartheta_{11} \vartheta_{11} \vartheta_{11} \vartheta_{11} & = 2 \vartheta_{11} \vartheta_{11} \vartheta_{11} \vartheta_{11} \, , \\ \nonumber
(R9): \vartheta_{00} \vartheta_{00} \vartheta_{01} \vartheta_{01} - \vartheta_{01} \vartheta_{01} \vartheta_{00} \vartheta_{00} - \vartheta_{10} \vartheta_{10} \vartheta_{11} \vartheta_{11} +\vartheta_{11} \vartheta_{11} \vartheta_{10} \vartheta_{10} &=  -2 \vartheta_{10} \vartheta_{10} \vartheta_{11} \vartheta_{11}  \, , \\ \nonumber
(R11): \vartheta_{00} \vartheta_{00} \vartheta_{10} \vartheta_{10} + \vartheta_{01} \vartheta_{01} \vartheta_{11} \vartheta_{11} - \vartheta_{10} \vartheta_{10} \vartheta_{00} \vartheta_{00} - \vartheta_{11} \vartheta_{11} \vartheta_{01} \vartheta_{01} &= 2 \vartheta_{01} \vartheta_{01} \vartheta_{11} \vartheta_{11} \, , \\
(R15): \vartheta_{00} \vartheta_{00} \vartheta_{11} \vartheta_{11} + \vartheta_{01} \vartheta_{01} \vartheta_{10} \vartheta_{10} - \vartheta_{10} \vartheta_{10} \vartheta_{01} \vartheta_{01} -\vartheta_{11} \vartheta_{11} \vartheta_{00} \vartheta_{00} &= 2 \vartheta_{00} \vartheta_{00} \vartheta_{11} \vartheta_{11} \, .
\end{align}
In the above the arguments of the theta functions on the left hand side are $x,y,u,v$, in that order, and on the right hand
side the arguments are $x_1,y_1,u_1,v_1$, in that order. 

When $x=y=u=v=0$, using the fact that $\vartheta_{11}(0)=0$, (R5) reads
\be
 \vartheta_3^4 - \vartheta_4^4- \vartheta_2^4 = 0
 \ee
 which  is Jacobi's ``abstruse identity" demonstrating equal numbers of space-time bosons and fermions in the GSO projected
 superstring.

\end{document}